\documentclass{sig-alternate-2013}
\usepackage{url}
\usepackage{multirow}
\usepackage{graphicx}

\newfont{\mycrnotice}{ptmr8t at 7pt}
\newfont{\myconfname}{ptmri8t at 7pt}
\let\crnotice\mycrnotice%
\let\confname\myconfname%

\permission{Permission to make digital or hard copies of all or part of this work for personal or classroom use is granted without fee provided that copies are not made or distributed for profit or commercial advantage and that copies bear this notice and the full citation on the first page. Copyrights for components of this work owned by others than ACM must be honored. Abstracting with credit is permitted. To copy otherwise, or republish, to post on servers or to redistribute to lists, requires prior specific permission and/or a fee. Request permissions from Permissions@acm.org.
}

\begin{document}
%
 \conferenceinfo{COSN'14,}{October 1--2, 2014, Dublin, Ireland.}
 \copyrightetc{Copyright 2014 ACM \the\acmcopyr}
 \crdata{978-1-4503-3198-2/14/10\ ...\$15.00.\\http://dx.doi.org/10.1145/2660460.2660470\\}

\title{Online Privacy as a Collective Phenomenon}

\numberofauthors{3} 
%
\author{
%
%
\alignauthor
Emre Sarigol\\
       \affaddr{ETH Zurich}\\
       \affaddr{Weinbergstrasse 56/58}\\
       \affaddr{Zurich, Switzerland}\\
       \email{semre@ethz.ch}
\alignauthor
David Garcia\\
       \affaddr{ETH Zurich}\\
       \affaddr{Weinbergstrasse 56/58}\\
       \affaddr{Zurich, Switzerland}\\
       \email{dgarcia@ethz.ch}
\alignauthor 
Frank Schweitzer\\
       \affaddr{ETH Zurich}\\
       \affaddr{Weinbergstrasse 56/58}\\
       \affaddr{Zurich, Switzerland}\\
       \email{fschweitzer@ethz.ch}
}


\maketitle

\begin{abstract}
The problem of online privacy is often reduced to
individual decisions to hide or reveal personal information in online social networks (OSNs). However, with the increasing use of OSNs, it becomes more important to understand the role of the social network in disclosing personal information that a user has not revealed voluntarily: How much of our private information do our friends disclose
about us, and how much of our privacy is lost simply because of online social
interaction? Without strong technical effort, an OSN may be able to exploit the assortativity of human private
features, this way constructing shadow profiles with information that users
chose not to share. Furthermore, because many users share their phone and
email contact lists, this allows an OSN to create full shadow profiles for people who do not even have an account for this OSN.

We empirically test the feasibility of constructing shadow profiles of
sexual orientation for users and non-users, using data from more than
3 Million accounts of a single OSN. We quantify a
lower bound for the predictive power derived from the social network of a user, to demonstrate
how the predictability of sexual orientation increases with the size
of this network and the tendency to share personal information. This allows us to define a \emph{privacy leak factor} that
links individual privacy loss with the decision of other individuals to disclose information.
Our statistical analysis reveals that some individuals are at a higher
risk of privacy loss, as prediction accuracy increases for users with a
larger and more homogeneous first- and second-order neighborhood of their social network.
While we do not provide evidence that shadow profiles exist at all,
our results show that disclosing of private information is not restricted to an individual
choice, but becomes a collective decision that has implications for policy and
privacy regulation.
\end{abstract}

\newpage


\section{Introduction}

Our society is increasingly grounded on information and
communication technologies, in which protecting one's privacy might not be an
individual choice \cite{Castells2000}.  In online social networks (OSNs), the
characteristics of each user is determined primarily by its
connections, rather than by its intrinsic properties. Hence, from an
individual's perspective, isolation is often not a desirable option
\cite{Stalder:2002}. To that end, the issue of protecting one's privacy within the
OSN relates largely to the community the individual is embedded in,
and how it is handled, if at all, by the community at large.

Although the existence of mass surveillance and the imminent threats it poses are known by many, studies in a number of fields show that people do little to protect their privacy against surveillance \cite{Stalder:2002}. In an OSN, users are often incentivised to share personal data, e.g. by offering some sort of benefit or personalization as a service (e.g. recommender
systems). But incentives also arise from social influences, e.g. from social surveillance of peers to receive attention and to
reinforce existing relationships 
\cite{Marwick2012}. When an OSN provider has access to the
contacts of users, it gains stronger predictive power. Together with the content willingly produced by users, there are ways to extract probabilistic
profiles of other users, even about persons who did not have an account in the given OSN \cite{Boyd2012}. 

On the aggregated level, this leads to an imbalance between the knowledge that a single user has about the OSN provider and the knowledge that the provider has, or is able to deduce, about individual users and even about persons that are not users. There is no way of knowing how the information provided by an everyday user to the OSN can be utilized, and there are no clear policies about this either. The usage often stretches
from personalization to social discrimination, without the user's
knowledge \cite{lyon2002surveillance}. 
A \texttt{Facebook} bug revealed in 2013 is an appropriate example for one of the many ways the information provided by users may be utilized. 
According to the reported bug, \texttt{Facebook} attempted to obtain
users' off-site email addresses and phone numbers, gathered from the
contact lists shared by other users. It appeared that the covertly
collected information was then being stored in each \texttt{Facebook} user's
invisible \textit{ Shadow Profile } that is somehow attached to
accounts \cite{Article2}. Digital trends \cite{Article1} defines a
\texttt{Facebook} Shadow Profile as \textit{"a file that \texttt{Facebook} keeps on you
  containing data it pulls up from looking at the information that a
  user's friends voluntarily provide. You're not supposed to see it,
  or even know it exists."}  \texttt{Facebook} reacted to the incident to fix
the leak as soon as possible. However, what has remained not fixed until now is
their policy. In a set of interviews, \texttt{Facebook} officials claimed that
obtaining third party user data on individuals in this manner was not
a privacy breach since the data has been submitted voluntarily by
members of \texttt{Facebook}, which make the data a property of \texttt{Facebook}
\cite{Article3}. This argument is backed up by the following statement
in \texttt{Facebook} Terms of Service:

\begin{flushleft}
\textit{We receive information about you from your friends and others,
  such as when they upload your contact information, post a photo of
  you, tag you in a photo or status update, or at a location, or add
  you to a group. When people use \texttt{Facebook}, they may store and share
  information about you and others that they have, such as when they
  upload and manage their invites and contacts }\cite{FBToS}. 
\end{flushleft}

\texttt{Facebook for Mobile} alone has over one billion
users that agreed to their terms of service, which allows the
application to \emph{"read data about your contacts stored on your
  phone, including the frequency with which you've called, emailed, or
  communicated in other ways with specific individuals"}.  It is not difficult to imagine the massive amount of 
ongoing data acquisition based on such intact privacy policies.
Therefore, it becomes an imminent question to what extent can an OSN be turned into a tool that acquires data to profile the whole society, just because some individuals have become members of
that OSN.  Our main contribution to this discussion is to demonstrate to what extent the information of an OSN provider about its users can be used to quantify knowledge about the individuals of our society, at large. We use an empirical
dataset from \texttt{Friendster}, a large online social networking
site that preceded \texttt{Facebook}. This dataset, which is publicly
accessible in the Internet
Archive\footnote{\url{https://archive.org/details/archive-team-friendster}},
allowed us to evaluate the power that Friendster had to create shadow
profiles. 

Our aim is not to provide new tools or algorithms to improve the
accuracy of the knowledge that an OSN provider can
possess. Instead, we aim to apply state of the art statistical analyses and
machine learning techniques to quantify the extent to which individual
privacy is leaked by the activity of others in an OSN, and
to empirically test how the individual decision to reveal 
information turns into a collective phenomenon to disclose privacy. 
In our analysis, we study two interrelated problems. First we explore the
\textit{Partial Shadow Profile} problem, in which an OSN infers private information that its users chose not to
share. Second, we address the \textit{Full Shadow Profile} problem, in which an OSN provider discovers private information about individuals who do not even have an account there, solely based on personal information and
contact lists shared by its actual users.

In this work, we focus on sexual orientation as a relevant and sensitive
private information the disclosure of which should be in control of the users. The
combination of gender and sexual orientation creates a set of classes
that appear with inhomogeneous frequencies, which is often the case in
real-life prediction problems of different domains (e.g. political
affiliation). For each user, we construct a simple social context 
based on frequency measures on the neighborhood at increasing
distances. 
We quantify privacy leak factors for different sexual
orientation groups and analyze how they are affected by two main factors,
the network size and the disclosure parameter, i.e. the 
ratio of users sharing their contact lists and/or private information
with the OSN. We further analyze how the coefficients of larger (i.e. majority) and smaller (i.e. minority) sexual orientation groups compare with respect to these two factors.

\section{Related Work}
\label{sec:RelatedWork}

Understanding privacy in OSNs starts with the individual motivation to
share personal information and its associated risk of sharing this
information with undesired contacts
\cite{Acquisti2005,Johnson2012}. Most OSNs include highly customizable
modules to control privacy settings, which can lead to higher efforts
and uncertainty how to use the site \cite{Liu2011}, or to distancing
from those users that have a lower awareness of possible data leakage \cite{Malandrino2013, Krishnamurthy2010, Brandimarte2013}.  Recent technologies promise
to alleviate user privacy concerns.  For example, distributed
recommender systems can put a limit to privacy disclosure \cite{Isaacman2011}, deployment of OSNs in the cloud can avoid the centralization of user
data \cite{Wilson2011, Zhang2013}, techniques for picture encryption
\cite{Tierney2013} and content anonymization \cite{Puttaswamy2009} can
prevent undesired access to private content.

Private information about users can be a source of wealth, e.g. by significantly increasing the
revenue of personalized advertisement \cite{Gill2013}. This creates incentives
for OSNs to  share private user information with third parties, from which the user does not necessarily benefit. A possible solution for this dilemma is to create
monetization schemes that allow users to set up the price they request
for companies to access their private information \cite{Riederer2011},
effectively creating \emph{privacy butlers} that automatically control
 privacy \cite{Wishart2010, Lanier2013}. This
approach can be criticized for its ethics about the value of privacy, asking
if market dynamics would push less wealthy individuals to have no
privacy \cite{Morozov2013}. Additionally, the monetization of privacy
relies on systems that would allow an individual to have full access control
of its privacy, which is hardly realistic. 

Even with full individual control, the possibility of
third-parties to infer private attributes still exists \cite{Mondal2014}.
The discovery of unknown/hidden parts of a network based on its visible
properties is a well
studied problem, in particular with respect to 
\textit{link prediction}
\cite{ErdosGT12,KimL11,Liben-Nowell:2003:LPP:956863.956972}. Such hidden links have been shown to be predictable by
geographic coincidences \cite{Crandall08122010}, using geotagged photo
data from Flickr. The method introduced in \cite{Crandall08122010}
utilizes the number and proximity in time and space of co-occurences
among pairs of individuals to infer the likelihood of a social tie
between them.  The link prediction problem has also been applied to
predict links between non-users of \texttt{Facebook} \cite{Horvat:2012jk},
given only the link information towards
non-members from the known network.  Additionally, the \textit{network completion}
problem aims to infer both missing links and nodes, where it 
has been shown that the missing part of the network can be inferred based
only on the connectivity patterns of the observed part \cite{KimL11}.

Previous studies of user privacy have focused on \textit{sensitive
  attribute inference} problems, where  user private attributes are detected 
based on a mix of public profiles in the network, friendship links and
group membership information of private users
\cite{Zheleva:2009:JJI:1526709.1526781}. Specifically,  within the \textit{friendship identification and inference}
attack \cite{JinLJ13},
a user might aim to infer private attributes of another user. Given that the attacker and the target are
direct or 2-distant neighbors, the success of such attacks depends on
network topological properties, such as the position of the attacker
in the network. Furthermore, iterative algorithms can effectively
label nodes by propagating information to their neighborhoods
\cite{GayoAvello:2011:LDY:1995966.1995991}. A wide variety of models
have aimed at predicting different private features, such as gender,
age, political orientation \cite{Rao:2010:CLU:1871985.1871993}, home
location \cite{Pontes2012}, and academic profiles \cite{Mislove2010}.
In the context of sexual and romantic relationships, two previous
works are especially relevant. First, the \emph{``gaydar''} experiment
\cite{JerniganM09} showed that homosexual male users can be detected
based on the amount of friends of the same type they had on
\texttt{Facebook}. Second, a recent article \cite{DBLP:BackstromK14} proposes a
new measure of \textit{dispersion} and applies it to a
large \texttt{Facebook} dataset in order to predict which of a user's friends
is their romantic partner.


In this article, we evaluate the accuracy
of partial and full shadow profiles for the sexual orientation of
users and non-users of the \texttt{Friendster} social network. Our
analysis builds on the sequence of users joining \texttt{Friendster} to
evaluate predictions over individuals without a user account in a
similar manner as done in \cite{Horvat:2012jk} where the links between
non-users are inferred. Knowing in which sequence the users joined \texttt{Friendster} has freed us from having to utilize a network growth model in our analysis. Furthermore, we pay special attention to the ratios of friends belonging to each orientation in the neighborhood of users at a given time in the growth of the network. Our
results should be compared with previous work on sexual
orientation of users in smaller datasets \cite{JerniganM09}.  To our
knowledge, our work is the first to address the possibility of
creating full shadow profiles for the sexual orientation of non-users
from a large scale OSN.



\section{Data Description}
\label{sec:dataDescription}

Before its social networking functionalities were discontinued,
\texttt{Friendster} was crawled by the \texttt{Internet Archive}, leaving a snapshot of
all the publicly available information at that moment.  Our previous
analysis of the connectivity patterns of the network \cite{Garcia2013}
reveals that the first 20 Million users of \texttt{Friendster} were largely
located in the US, before the OSN spread to other countries.  The growth of
\texttt{Friendster} in the US stopped because of the competition with \texttt{MySpace} and
\texttt{Facebook} \cite{Ribeiro2014}. This allows us to analyze these
initial 20 Million users as a subset of US users of the OSN. 

The amount of information about each user available in the \texttt{Friendster}
dataset depends on the privacy settings of the user. Most of them
allowed their friendship lists to be publicly available, and some of them
also let other users to see private features explicitly given by the user,
such as age and gender.  
Within the subset we considered, 3,431,335 users had public profiles which were captured by the crawl, including the personal information explained in Table \ref{tab:ProfileDesc}. This subset contains a total of 11,074,009 undirected friendship links among these public profiles only, resulting in an average degree of 3.23. 

\begin{table}[htbp]
\centering
\small
\begin{tabular}{|p{1.5cm}|p{6cm}|}
\hline Feature & Description \\ \hline 
User ID & integer \\ 
Name & string \\ Birth date &
date \\ Gender & \emph{Male}, \emph{Female}, or \emph{Unspecified} \\ \hline
Interests$^{*}$ & \emph{Friends}, \emph{Activity Partners}, \emph{Just looking around}, \emph{Fans},
  \emph{Dating Women}, \emph{Relationship with Women}, \emph{Dating Men}, \emph{Relationship with
  Men}, \emph{Dating Men and Women}, \emph{Relationship with Men and
    Women} \\ \hline
Relationship status & \emph{Single}, \emph{Married}, \emph{In a Relationship}, \emph{Domestic
Partners}, \emph{It's Complicated} \\ \hline
\end{tabular}
\caption{\texttt{Friendster} public profile features. The
    interests feature contains one or more of its possible
    values. \label{tab:ProfileDesc}}
\end{table}

In addition, each user has an id number that indicates the order in
which the user joined the social network, allowing us to construct
time-dependent vectors of feature distributions in user neighborhoods,
as described in the following section.

\begin{figure}[htbp]
  \centering
    \hspace*{-0.5cm} 
  \includegraphics[width=1.15\linewidth]{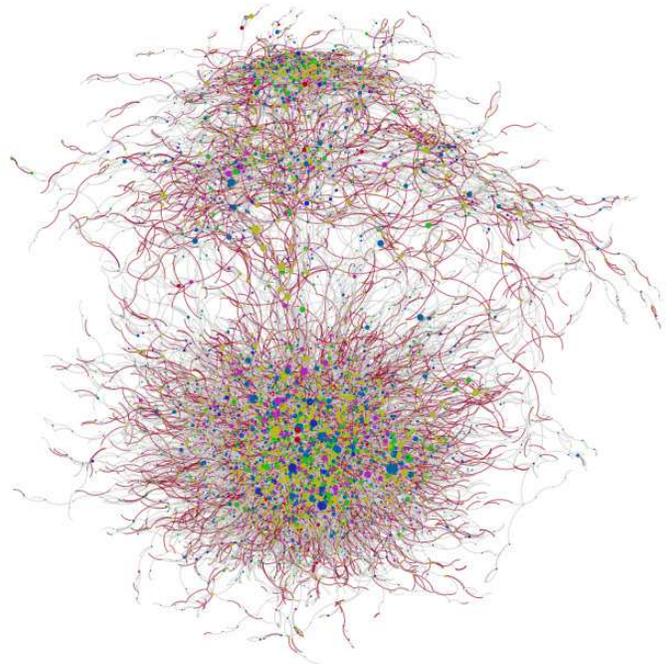}
  \caption{\textbf{The network for a subset of Friendster users. The red edges represent assortativity, where the endpoint nodes are in the same sexual orientation class. The node colors correspond to the sexual orientation class.}}
  \label{fig:FriendsterNetwork}
\end{figure}

The network in Figure \ref{fig:FriendsterNetwork} displays the \texttt{Friendster} network among a randomly selected $10\%$ of the users with public profiles, where node colors represent the sexual orientation class, and the colored edges represent assortativity where two endpoints share the same sexual orientation class. About $30\%$ of all the edges are assortative in this representative network, which suggests that it is common to form links based on sexual orientation.

\section{Methods}
\label{sec:Methods}

Being aware that sexual orientation and gender can be defined in a
variety of ways \cite{Boyd2001}, we use the simplest classification
available in our data: gender as birth sex (male or female) and sexual
orientation simplified to the set of possible combinations of interest
towards the two genders. This way, we combine gender with the explicit romantic
interest in other genders, as specified by \emph{Dating} and
\emph{Relationship} interests towards different genders introduced in Table
\ref{tab:ProfileDesc}. Each user can be assigned to one of the eight classes
with respect to their sexual orientations, which are described in
Table \ref{tab:UserOrient}. Additional features of sexual orientation
can capture other activities, group identities, or political
standpoints, and other features that cannot be empirical measured in
our dataset.

\begin{table}[!t]
\centering
\small
  \hspace*{-0.4cm} 
\begin{tabular}{|c|c|c|l|c|c|}
\hline
Gender  & M  &  F  & Class   & Label & $\%$ users \\ \hline
Female 	& No & No	& Female without interest &\textbf{FF} 	& 28.2  \\ 
Male 	& No & No	& Male without interest & \textbf{FM} 	& 26.4  \\ 
Female 	& Yes & No	& Heterosexual female & \textbf{HeF}	& 9.3 \\ 
Male 	& Yes & No	& Homosexual male & \textbf{HoM}		& 1.9 \\ 
Female 	& No	& Yes	& Homosexual female & \textbf{HoF} 	& 1.0 \\ 
Male 	& No & Yes	& Heterosexual male & \textbf{HeM}		& 19.9 \\ 
Female 	& Yes & Yes	&  Bisexual female & \textbf{BiF}		& 6.8 \\ 
Male 	& Yes & Yes	&  Bisexual male & \textbf{BiM}			& 6.5 \\ 
\hline
\end{tabular}
\caption{User orientation classification. An "interested
    in" relationship may stand for interested in dating or having a
    relationship.  \label{tab:UserOrient}}
\end{table}

\begin{table}[htbp]
\centering
\begin{tabular}{|c|p{6cm}|}
\hline Features & Description \\ \hline 
Profile & Age, gender, relationship status, Sexual
orientation \\ \hline
$n_k$ & Number of users at distance $k$ \\
$a_k$ & Average age of friends at distance $k$\\
$g_k$ & Gender counts at distance $k$\\
$r_k$ & Relationship counts at distance $k$\\
$i_k$ & Romantic interest counts at distance $k$\\
$x_k$ & Sexual orientation counts at distance $k$\\
$x_w$ & Weighted frequency of friends of each sexual orientation\\
\hline
\end{tabular}
\caption{Features of the user vector. Neighborhood frequencies
    are computed for distances 1, 2, and 3, for each possible value of
    the profile features. \label{tab:Vector}}
\end{table}

For each user in the dataset, we built a feature vector including their
profile information, and different metrics of the distribution of
features in their neighborhood at distances up to 3. For each distance
$k$, we calculated the amount of users at that exact distance ($n_k$),
and within those users, we counted the amounts of users with each
possible value of gender, relationship status, romantic interest, and
sexual orientation. To measure age in the neighborhood, we computed
the average age of the users at distance $k$ ($a_k$). Since previous
research suggests that the most indicative factor is the sexual
orientation of the first neighbors of the user \cite{JerniganM09}, we
computed an additional weighted count of friends of each sexual
orientation, weighting each link by the amount of common friends that
the two users have. 
The features of this vector are summarized in Table \ref{tab:Vector}.





\newpage

\section{Partial Shadow Profiles}
\label{sec:PartialShadows}

Our first step was to explore the \emph{partial shadow profiles} problem.  We
define partial shadow profiles as enhanced data of an OSN provider about
its users, covering personal information these users did not initially
agree to share. We test the OSN provider's ability to construct
partial shadow profiles over the set of users that have initially
disclosed their romantic interest towards at least one gender, leaving out
users of the classes FF and FM. This leaves us with 1,027,400 users and six classes, with feature vectors built
over the network including all 3.3 Million users to reach
neighborhoods at larger distances.

We arrange the data of these users as follows: We choose a \textit{
  partial disclosure parameter}  $R \in \{ x/10 : x \in \mathbb{N}, x < 10 \} $,
  defined as the
probability that a user has shared sexual interest information with
the social network. For a given $R$, we include users in the
\emph{training} set with probability $R$, and leave them for the
\emph{test} set with probability $1-R$. 
The training set contains those users whose sexual orientation class are known, and the test set contains those users whose class or other user features (e.g. gender and age) are hidden. This reproduces the problem setup that
the OSN provider faces when constructing partial profiles: a set of
its users have disclosed their orientation, but others chose not
to. We preserve the friendship links of all users, including those
in the \emph{test} set 
and build the user vector for the \emph{training} set of users, using all the links and only the user features within the \emph{training} set, since the user features of the \emph{test} set of users are hidden. We use this vector to train a Random Forest Classifier, and use the resulting classifier to predict
the sexual orientation of users in the \emph{test} set. Since both the
training and the test cases are randomly chosen, we repeat this 10 times for
each $R$. 

Through these 10 repetitions, we aim at understanding the dependence between 
the tendency to share personal information of the users of an OSN, and
the predictability of the sexual orientation of those users who chose
not to share that information. In particular we want to understand under which conditions this prediction would outperform a random estimator and by what factor.

\subsection{Prediction Results}

\begin{figure}[t!]
  \centering
  \includegraphics[width=1.0\linewidth]{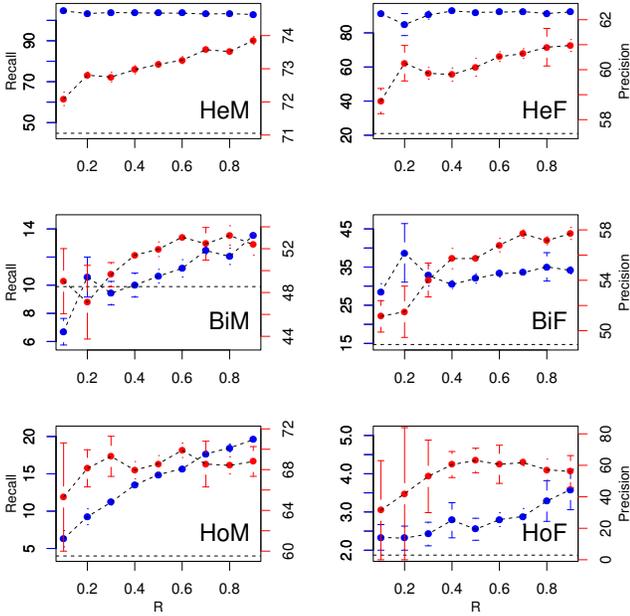}
    \caption{Recall and Precision (\%) for each class versus $R$ for the
      partial shadow profile problem. The blue lines and the left
      y-axis show the recall values, whereas the red lines and the
      right y-axis show the precision values as $R$, the partial
      disclosure parameter, grows. The dashed black line shows the
      base rate, the percentage of users for each class within the
      whole data set.    \label{fig:PrecRecallPartial}}
\end{figure}

For each value of $R$ and random samplings of training and test users, we
computed the \emph{Precision} and \emph{Recall} values for each of the six
sexual orientation classes.  Figure \ref{fig:PrecRecallPartial} shows
the mean values over the 10 runs of each value of $R$.

We observe that for all classes, \emph{Recall} can reach values much higher
than the base rate, which is equivalent to the percentage of users that belong
to each class.  This holds for low values of $R$ for all classes but
homosexual females and bisexual males, which require $R>0.3$ to have a
precision above the base rate.  For the case of homosexual females,
which constitutes $2\%$ of all users, the \emph{Precision} increases up to
$60\%$ but the \emph{Recall} values increase marginally between $2\%$ and
$4\%$, showing that some homosexual females can be detected with high
\emph{Precision}, but the vast majority of them cannot be predicted by the
Random Forest Classifier. The most striking results are for
homosexual males, where both \emph{Precision} and \emph{Recall} are several times
above the base rate, and for the majority classes of heterosexual
males and females, which also show large values of \emph{Precision} and
\emph{Recall}.

\emph{Precision} values alone indicate that the accuracy of predictions
increases significantly with higher values of $R$. To empirically test
the relationship between prediction results and the partial disclosure
parameter, we computed Cohen's Kappa Coefficient \cite{Cohen1968} for
each class and classification run:
\begin{equation}
\kappa = \frac{Pr(a) - Pr(e)}{1 - Pr(e)} 
\end{equation}
where $Pr(a)$ is the relative observed agreement between
classification and test data, and $Pr(e)$ is the hypothetical
probability of chance agreement, using the observed data to calculate
the probabilities of each observer randomly saying each category. 

If the raters are in complete agreement then $\kappa = 1$. If there is no
agreement among the raters other than what would be expected by chance
(as defined by $Pr(e)$), $\kappa = 0$. Cohen's Kappa $\kappa$ captures
a combination of \emph{Precision} and \emph{Recall} similar to the $F_1$ value,
but includes the comparison to the baseline of a random classifier in
its calculation. Thus, the performance of a random classifier would
tend towards $\kappa=0$, while the value of $F_1$ would depend on the
distribution of classes in the dataset.

\begin{figure}[htbp]
\centering
    \includegraphics[width=\linewidth]{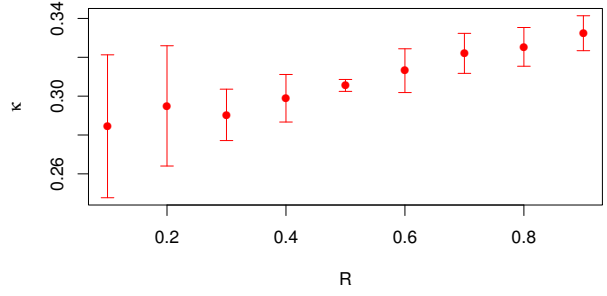}
    \caption{Cohen's Kappa coefficient versus $R$ for the partial
      shadow profile problem.  \label{fig:kappaPartial}}
\end{figure}

Since Kappa's coefficient is independent of class size and thus is
resilient to biases introduced by differing class sizes, we aggregate
the performance of the classifier for individual classes into a single
average $\kappa$. Figure \ref{fig:kappaPartial} displays the average
Cohen's Kappa coefficient over all classes.

We observe that as the partial disclosure parameter $R$ grows, an OSN
provider would be able to predict, with higher accuracy, the sexual
orientation of those users that did not share it. We statistically
test this observation through a \textit{privacy leak factor}
for each user class, computed as the weight of $R$ in a linear
regressor with an intercept and $\kappa$ as dependent variable.  Table
\ref{tab:PartialLeakingFactors} shows the statistical results for each
of the six fits, as well as the fit for the privacy leak factor
computed for all classes on the average $\kappa$.


\begin{table} [htbp]
  \centering
  \begin{tabular} {|c|c|c|c|}
  \hline
  Class           & Class $\%$  & priv. leak. factor   & $p$-value   \\ \hline
  \textbf{HeM}  & 47          &  0.02           & $2.9 \times 10^{-7}$  \\
  \textbf{HeF}  & 23          &  0.04           & $2.1 \times 10^{-3}$  \\
  \textbf{BiM}  & 9           & 0.09            & $4.9 \times 10^{-4}$  \\
  \textbf{BiF}  & 16          & 0.06            & $5.2 \times 10^{-2}$  \\
  \textbf{HoM}  & 4           & \textbf{0.24} & $7.1 \times 10^{-6}$  \\
  \textbf{HoF}  & 2           & 0.02            & $8.6 \times 10^{-3}$  \\
  \hline        
  \textbf{$\overline{\kappa}$} & - & 0.12  & $1.1 \times 10^{-3}$ \\
  \hline
  \end{tabular}
  \caption{privacy leak factors for each class and average
    $\kappa$ in the partial shadow profiles problem. All estimates are
    below the $0.01$ significance level, with the exception of
    \textbf{BiF}. \label{tab:PartialLeakingFactors}}
\end{table}

This statistical analysis demonstrates that all classes have a significant and
positive privacy leak factors, with the exception of bisexual
females, for which the p-value was above $0.01$. The size estimates of
the privacy leak factor differ across classes, having
relatively low values for heterosexual male and female, and for the
homosexual female class. The largest values are present for the
homosexual and bisexual male classes.  Homosexual males have low
predictability under $R=0.1$, since they constitute about $4\%$ of all
users, but the privacy leak factor is much higher than for other
classes, being estimated as $0.24$, 12 times larger than for the
largest class of heterosexual males.  This suggests that homosexual
male users that do not disclose their sexual orientation are at a
larger risk of privacy leakage if the tendency of other
users to share their sexual orientation becomes stronger.

Finally, the standard errors for the average $\kappa$, and the \emph{Precision} and \emph{Recall} of individual classes reveal that, across the 10 runs for each $R$, the prediction accuracy does not vary much for a given $R$, especially for higher values of $R$. This suggests that the prediction accuracy does not rely on which users of the OSN have revealed their sexual orientation, given that large enough ($R\simeq0.3$) percentage of the population share personal information.

\newpage

\section{Full Shadow Profiles}
\label{sec:FullShadows}

Full shadow profiles are the profiles that an OSN provider can generate about individuals  that do not have an account for this OSN. The idea is, when a user shares its contact list with the OSN, the provider can find out which email addresses do not have an  associated account and can generate a full shadow profile for these non-users. If those non-users appear in many contact  lists of OSN users, data mining techniques can be used to infer the home location, age, gender, etc, of the non-users. 

\begin{figure}[htbp]
  \centering
  \includegraphics[width=1.0\linewidth,trim={0 10cm 0 3mm},clip]{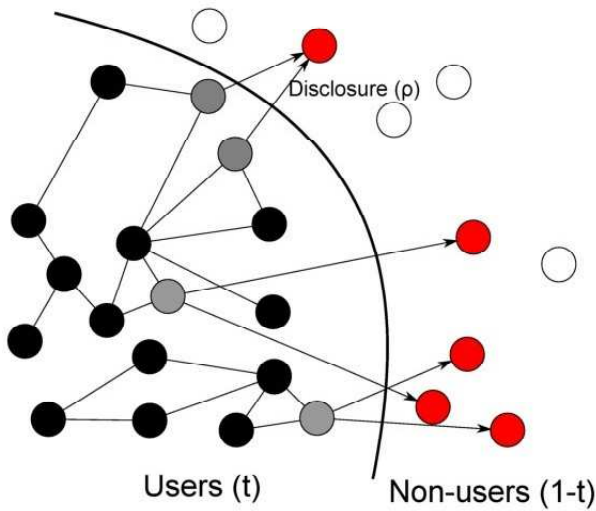}
  \caption{\small\textbf{Schema of the full shadow profile construction
      problem.}}
  \label{fig:ShadowProblem}
\end{figure}

For the Full Shadow Profiles problem, we arrange our data as follows: We select a parameter  $a \in \{ x/10 : x \in \mathbb{N}, x < 10 \} $, where each $a$ divides the whole user data of $N$ users into two sets:

\begin{itemize}
\renewcommand\labelitemi{--}
\small
\itemsep0em
	\item \textit{Inside} user set, which is of size $a \times N$ users
	\item \textit{Outside} user set, which is of size $(1-a) \times N$ users
\end{itemize}
In Figure \ref{fig:ShadowProblem}, the \textit{Inside} user set is represented by the combination of the black and gray nodes, and is the set of members of the OSN at time $t$. The \textit{Outside} user set is represented by the combination of red and white nodes, which are the set of users that are not part of the network at time $t$, hereafter denoted as  \textit{non-users}.

Furthermore, we introduce a \emph{disclosure parameter} $\rho \in \{ 0.5, 0.7, 0.9 \}$ which is the fraction of users in the \textit{Inside} user set that shared all of their contacts with \texttt{Friendster}. For \texttt{Facebook}, given the fact that every user of the \texttt{Face\-book for Mobile} initially has to agree that \texttt{Facebook} can access their contact list, $\rho$ would be closer to $1.0$.

Given a combination of $a$ and $\rho$, we measure: a) To what degree  is the OSN provider able to find out about the sexual orientation of the \textit{non-users};
b) How much is the social network confident about its findings about the \textit{non-users}, by measuring how much it can predict about its actual user base, i.e. \textit{Inside}.
To that end, we pursue the following method in three steps:

\begin{enumerate}
\itemsep0em
	\item We build a user vector for each user in the \textit{Inside} user set as described in Section \ref{sec:dataDescription} \footnote{For the full shadow profile analysis, we did not use 3-order neighborhood information due to the large number of simulations needed and computational limitations that were introduced}, discarding all their links to the non-users. 
	We then use this user vector to build a Random Forest classifier \cite{Breimann2001} over the user class of sexual orientation (hereby referred to as $RF_a$ ). 

	\item  For each user in the \textit{Inside} user set, we flip a biased coin where the outcome is heads with probability $\rho$ 
	and for those users that got tails, we discard all their links to the non-users. Using the remaining links, we build the user vector of the non-users who have at least one link to \textit{Inside}. This user vector represents the vector that the social network can construct for the non-users, using only the contacts of its users that shared their contact lists. 

	\item We use $RF_a$ to predict the sexual orientation of the non-users.

\end{enumerate}

While building the feature vectors for non-users in step 2, we discard all user attributes (relationship status, age and gender) from the feature vector, and keep only neighborhood information for each user. The reason we discard user attributes is to represent the real life situation where the social network knows nothing at all about the non-users at $t$. Since $RF_a$ will be used to classfy the resulting vector, $RF_a$ must also be built using the same features available in for the non-users. Therefore, we discard all user attributes also from the user feature vector in step 1.

Step 1 is run once for each $a$ where we acquire a corresponding $RF_a$ 
Step 2 and 3 are repeated 10 times for each $(a,\rho)$ pair, such that for each run, a different $\rho$ fraction of users share their contact lists, and hence a different set of links are preserved to the non-users.

\subsection{Prediction Results}

\begin{figure*}
    \centering
    \includegraphics[width=0.48\textwidth]{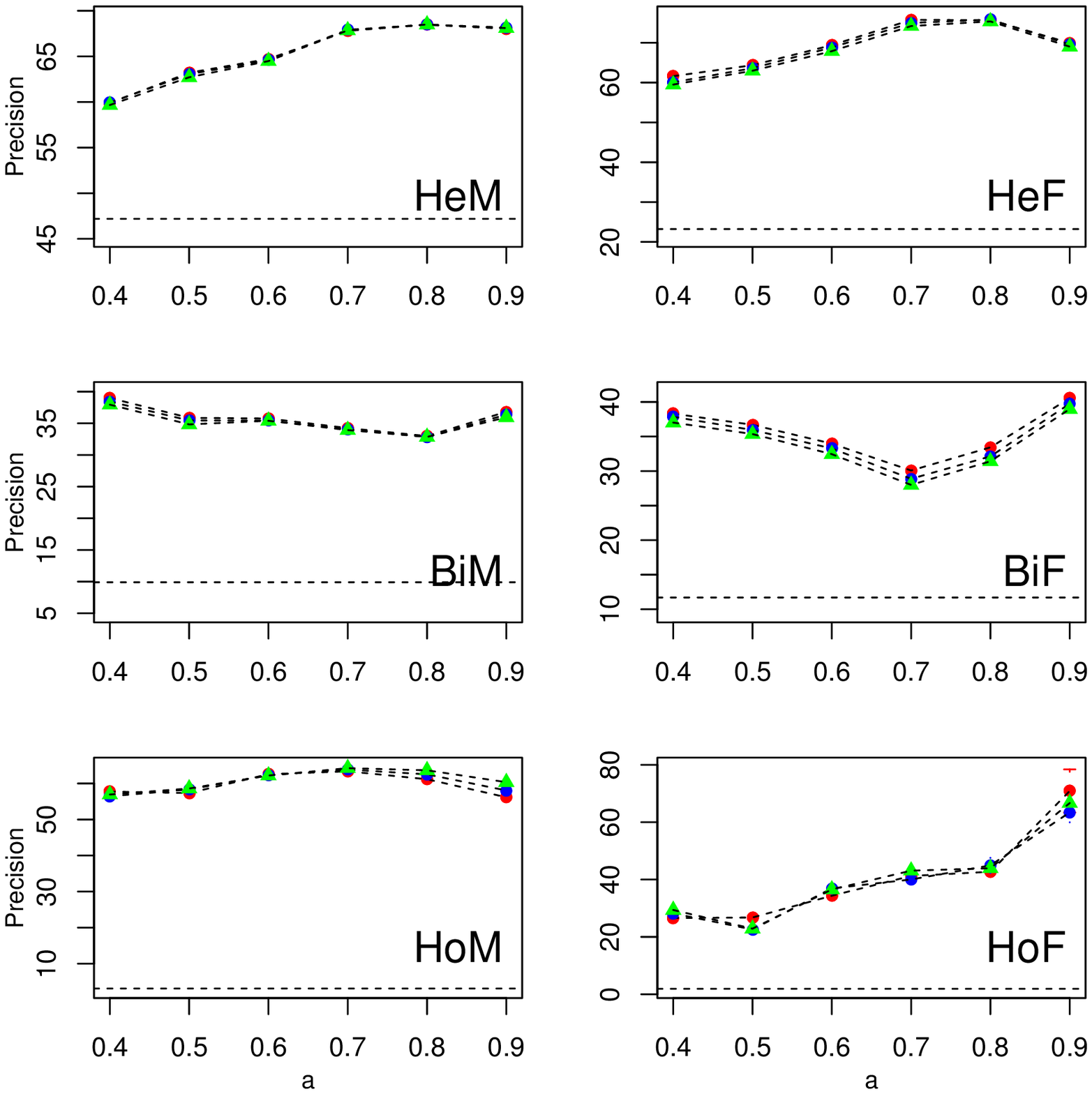} \hfill
    \includegraphics[width=0.48\textwidth]{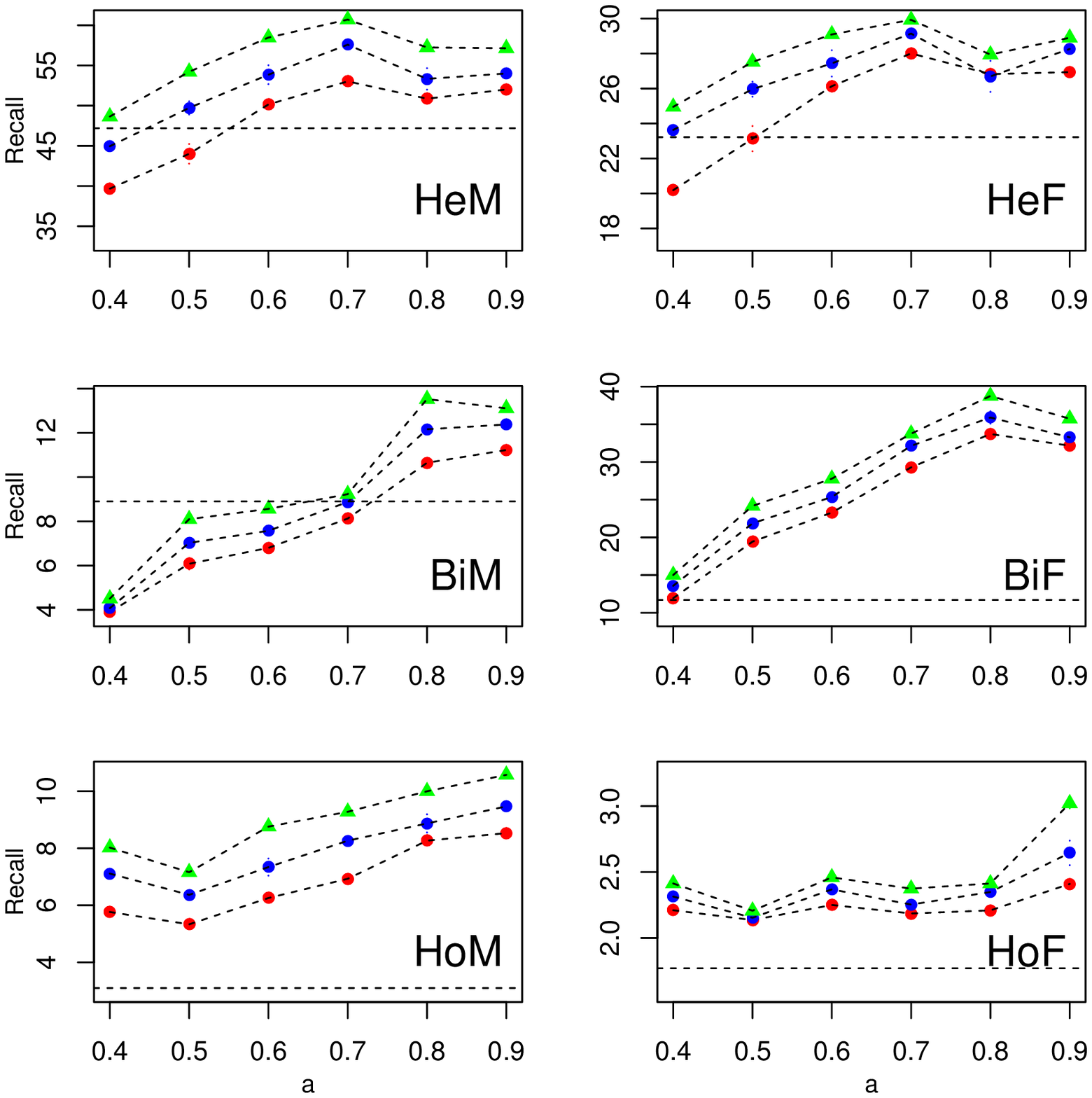}
  \caption{Precision and recall for each class in the full shadow
    profiles problem, for $\rho=0.5$ (red), $\rho=0.7$ (blue), and
    $\rho=0.9$ (green). The base rate of each class is given by the dashed black line. \label{fig:PrecRecallFull}}
\end{figure*}

In the full shadow profiles problem, non-users are subject of losing
privacy as other individuals join the OSN, potentially revealing their
contacts. We evaluate the performance of the $RF_a$
classifier over the set of non-users, for increasing values of $a$, to
test if prediction accuracy correlates with the size of the OSN. Since
these results are subject to increase with the disclosure parameter
$\rho$, we repeat the analysis for three different values of $\rho$.
We measured precision and recall over the complete set of non-users,
and report their mean values over 10 resamples in Figure
\ref{fig:PrecRecallFull}.

\begin{figure}[htbp]
    \centering
    \includegraphics[width=\linewidth]{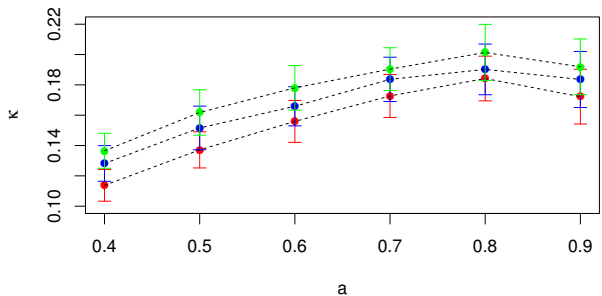}
    \caption{Cohen's Kappa in full shadow profiles for all classes
      versus $a$, for $\rho=0.5$ (red), $\rho=0.7$ (blue), and
      $\rho=0.9$ (green).  \label{fig:kappaFull}}
  \end{figure}

For all classes, there is an increasing trend in recall values with
$a$, as well as with $\rho$. $\rho$ plays a more significant role for recall than for precision. This is because as $\rho$ grows, feature vectors for more non-users can be constructed, leaving out less and less non-users from the predictions, which impacts recall over all non-users. Larger values of $a$ also contribute to
precision in most of the cases, but this increase seems negligible
compared to the distance between precision and base rate of each
class.  To further understand these trends, we computed Cohen's Kappa
for all classes over each evaluation. The average values of $\kappa$
versus $a$ are shown in Figure \ref{fig:kappaFull}, showing that the
predictive power of the classifier increases with $a$ and slightly
increases with $\rho$.

To statistically test for the presence of an increase of prediction
quality with $a$, we calculated privacy leak factors for the
full shadow profiles problem. In contrast with the partial shadow
profiles problem, where we linked the decision of users to disclose
their personal information, in the full shadow profiles problem we are
interested in knowing how the decision of some users to join the OSN
can influence the privacy of non-users. Thus, we compute privacy leak factor as the regression weight of $a$ in the $\kappa$ of the
classifier for different values of $\rho$. As shown in Table
\ref{tab:FullLeakingAll}, the privacy leak factor is positive
and significant for the three values of $\rho$, suggesting that an
overall privacy loss for non-users as the OSN grows.

\begin{table} [htbp]
  \centering
\begin{tabular}{|c|c|c|}
\hline
$\rho$  & privacy leak factor (all classes) & $p$-value  \\ \hline
0.5     & 0.12  & 0.007  \\
0.7     & 0.13  & 0.002   \\
0.9     & 0.16  & 0.002  \\
\hline
\end{tabular}
\caption{Privacy leak factor in full shadow profiles,
  calculated over $\kappa$ for all
  classes.  \label{tab:FullLeakingAll}}
\end{table}

The privacy leak factor for full shadow profiles is not
homogeneous for all sexual orientations. Table
\ref{tab:FullLeakingFactors} shows the privacy leak factors,
which are positive and significant for all classes but homosexual
female. The values of the privacy leak factor do not greatly
differ for the three values of $\rho$, suggesting that the main
driving factor of privacy loss of non-users is network growth.
Comparing across classes, the sexual orientation with the strongest
privacy leak factor is bisexual females, which had the least
significant counterpart for partial shadow profiles, as shown in Table
\ref{tab:PartialLeakingFactors}. This suggests that bisexual females
can be detected with higher accuracy when other users join the OSN,
rather than by disclosure of private attributes within the OSN.

\begin{table} [htbp]
  \centering \small
  \hspace*{-0.5cm} 
\begin{tabular} {|c|c|c c|c|c c|}
\hline
$\rho$  & Class  & priv.leak f. & $p$-value   & Class  & priv.leak f. & $p$-value  \\ \hline

0.5    &                 & 0.21  & 0.015 &  & 0.23  & 0.021    \\
0.7    & \textbf{HeM}    & 0.18 & 0.035 &   \textbf{HeF}  & 0.15  & 0.059   \\
0.9    &                 & 0.21   &  0.030 &               &  0.15&  0.051  \\
 \hline

0.5    &              &  0.17 & 0.00021 &                & 0.35 & 0.0012   \\
0.7    & \textbf{BiM} & 0.19  & 0.0011 &   \textbf{BiF}  & 0.32 & 0.0024   \\
0.9    &              &  0.21  & 0.0037 &               & 0.30 & 0.0055   \\
 \hline

0.5    &               & 0.11  & 0.029   &               & 0.0012 & 0.71     \\
0.7    & \textbf{HoM}  & 0.087 & 0.048    & \textbf{HoF}  & 0.0036 & 0.57     \\
0.9    &               & 0.097 &   0.046   &               & 0.0037 &   0.61    \\
\hline
\end{tabular}

\caption{Privacy leak factor in full shadow profiles for each
  class.  \label{tab:FullLeakingFactors}}
\end{table}

\subsection{Analyzing Prediction Results}
\label{sec:posthoc_analysis}

\begin{figure}[t!]
  \centering
    \hspace*{-0.3cm} 
  \includegraphics[width=1.0\linewidth]{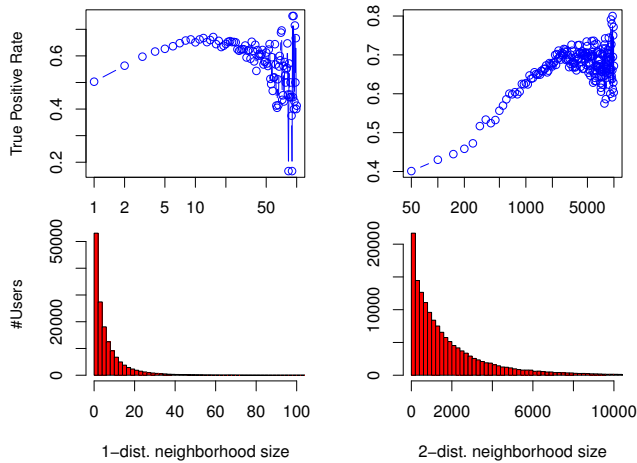}
  \caption{Bottom left: First order neighborhood size distribution
    where the sizes are the number of friends in the known part of the
    network who shared their contact lists. Top left: True Positive
    rate (TPR) for different first order neighborhood sizes where TPR
    is given by the ratio of number of correctly predicted users to
    the number of users that fall into each neighborhood size
    range. Bottom right: Second order neighborhood size
    distribution. Top right: TPR for different second order
    neighborhood sizes. All figures are derived from 10 simulations
    where $a=0.6$ and $\rho=0.9$.  \label{fig:FriendCounts}}
\end{figure}

We analyzed the properties that correctly predicted non-users have in
common, in order to shed light to other factors that may have played a
role in predictions. Figure \ref{fig:FriendCounts} shows the
distribution of the first order neighborhood size in the
\textit{Inside} for all non-users at $a=0.6$ and $\rho=0.9$, and the
corresponding \emph{true positive rate} (TPR) for each neighborhood
size range, calculated as the ratio of correctly classified users over
all classified users.  Non-users with at least one friend in the
\textit{Inside} are more likely to be predicted correctly (TPR =
0.52). In addition, the TPR increases with neighborhood size. At first
sight, this would mean that one friend in the \textit{Inside} set
gives the OSN provider enough power to accurately profile a
non-user. But looking at Figure \ref{fig:FriendCounts} (right column), which
shows the TPR distribution for different second order neighborhood
sizes of non-users, we observe a new dependence: Although the
second order neighborhood size is more heterogeneous, the TPR
increases significantly for larger sizes. Therefore, from a non-user's
perspective, not only the amount of friends in the OSN is a critical
factor, but how well connected those friends are.

\begin{figure}[t!]
  \centering
  \includegraphics[width=1.0\linewidth,keepaspectratio]{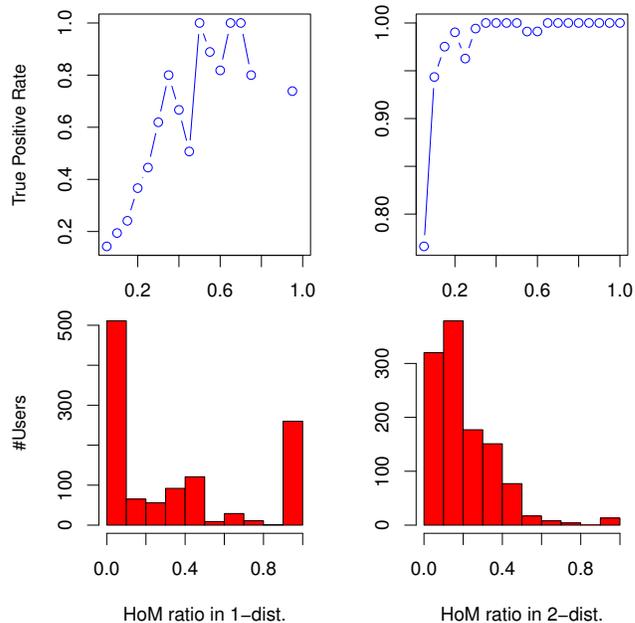}
  \caption{(bottom left) Distribution of HoM user ratios in the first
    order neighborhood of HoM non-users. (top left) TPR for HoM
    non-users in each HoM ratio range, where TPR is given by the ratio
    of number of correctly predicted HoM non-users to all HoM
    non-users. (bottom right) Distribution of HoM user ratios in the
    second order neighborhood of HoM non-users. (top right) TPR for
    HoM non-users in each second order HoM ratio range.  All figures
    are derived from 10 simulations where $a=0.6$ and
    $\rho=0.9$.  \label{fig:GayRatios}}
\end{figure}

We explored the assortativity of sexual orientation between users and
non-users, and how this can increase prediction accuracy.  As an
example, we look at homosexual male users, which is one of the smallest
classes. Figure \ref{fig:GayRatios} displays the distribution of
homosexual male user ratios in the first order neighborhood of
non-users of the same orientation, and the corresponding TPR for each
ratio of assortative links. It is more likely that homosexual male
non-users will be classified correctly as the ratio of links of the
same kind increases in their first order neighborhood, suggesting that
assortativity plays a significant role in privacy leakage. An
ratio of homosexual male friends of 0.1 is quite common, and displays
no significant affect on TPR (TPR = 0.09), although it is still larger
than the base rate. For a ratio of homosexual male friends between
$0.2$ and $0.5$, there is a clear increase in the TPR.

Noise is present between $0.5$ and $0.6$, due to the fact that there
are only very few non-users that fall in these categories. Figure
\ref{fig:GayRatios} suggests that the ratios of homosexual male
friends in the second order neighborhoods of non-users also correlate
with the respective TPR values, suggesting that higher order
assortative ties also influence privacy leakage.


These figures help us understand further what kind of dynamics contribute to privacy leakage in an OSN. The privacy leakage seems to be influenced by the size of the first order neighborhood, how many highly connected users exist within the first order neighborhood, and the assortativity across the first and second order neighborhoods with respect to the user's sexual orientation. The cross-dependence among these three factors seem to result in a large amount of privacy leakage. In this section we have looked at only a few of the possible factors and analyzed assortative relationships of only one sexual orientation class. The analysis can certainly be extended by looking into network features of known and unknown users and into different different classes.

\section{Discussion}
\label{sec:Discussion}



The privacy leak factor is not homogeneous for all sexual orientations. We showed that the privacy leak factors for large and small groups respond differently to changing network size and disclosure behavior. For example, the amount by which precision and recall digress away from their respective base rates for HoM individuals is much larger than and HeM individuals. Often, it is more risky for smaller groups to be compromised within the society, whereas larger groups are often not concerned by this risk. This is mostly due to peer pressure or to the fact that minority rights are legally not represented.

There are multiple factors that put individuals under privacy risk. We have shown that network size and disclosure parameter influence privacy risk for non-users. We have also suggested that this risk varies depending on how connected a non-user is to the network, and the nature of their connection (i.e. the homophilic nature of connections). 

The simulations are representative of a realistic network growth scenario and disclosure behavior. Our simulations are representative of a realistic network growth since we used \texttt{Friendster} user IDs, which are sequential with respect to the member's joining time, thus freeing our analysis to consult different growth models. Furthermore, the choice for the range of disclosure parameter,  $\rho \in [0.5, 0.7, 0.9]$, corresponds to a realistic scenario for the fraction of members sharing their contact lists. $\rho = 0.9$ is closer to reality as most people share their contact lists (either voluntarily or due to the accepted terms of use) when they subscribe to an online social network, as reports about \texttt{Facebook} reveal.
Finally, since we have made 10 runs for each  $a$ and $\rho$ pair, and since standard errors are quite marginal, we can conclude that the findings are rather representative of the conditions $a$ and $\rho$. 

Since the data of the first 20 Million \texttt{Friendster} users have been downloaded from the \texttt{Internet Archive}, the completeness of the dataset can only be guaranteed based on what the \texttt{Internet Archive} offers. Furthermore, we are bounded by about 3.3 Million of these users that disclosed any sort of sexual orientation, which results in a sparser network than that of all 20 Million users. Although the resulting network and user data is much bigger than the datasets that have been used in other works, a more comprehensive study can take into account all 20 Million profiles and study the 3.3Mil using all their links in the 20 Million. This would mean analyzing some 70 Million edges. However, given the number of user vectors computed for full shadow profiles alone where a user vector for the non-users is computed 10 times for each $(a,\rho)$ pair, resulting in 180 user vectors; and given the number of links that had to be traversed for computing each vector and the computational limitations introduced thereof, we believe that our results do provide a comprehensive analysis of privacy leakage. We have also not provided further analysis of factors that may play a role in prediction accuracy other than the ones discussed in Section \ref{sec:posthoc_analysis}. A more comprehensive analysis at this stage can answer the question which neighborhoods in larger distances still play a significant role in putting users under privacy leakage risk.





\section{Conclusions}

We presented an analysis of the social component of privacy, and how
the decisions of some users to disclose private information impacts the chances of other users to
maintain their privacy. This provides an indirect coupling between seemingly unrelated user decisions, as it was also observed in other 
online communities where the decision of some users to become incative influences other the activity of other users
\cite{Garcia2013}. Users in isolation face lower risks of losing
privacy than when they interact with each other. The same way as
social interaction leads to the emergence of conventions
\cite{Kooti2012}, it can also undermine the quality of collective
decisions \cite{Lorenz2011}, posing the question of how much private
information a user loses just for interacting with others. Our
work focused on sexual orientation, resonating within works on
gender-aligned interaction in online communities \cite{Garcia2014}.
But it keeps open to study privacy leakage in other kinds of
private information, such as age or marital status.

We showed that the privacy leak factors for large and small groups
respond differently to changing network size and disclosure
behavior. For example, privacy leak factors are higher for
homosexual males than for heterosexual males in the partial shadow
profiles problem, showing that the former group loses more privacy as
other users share their sexual orientation with the OSN provider.  We
have shown that network size and disclosure parameter influence
privacy risk for non-users in the full shadow profiles problem, and that this risk varies depending on how
connected a non-user is to the network, and the assortative nature of
their connections. This poses a simple conclusion: not having an
account in an OSN does not guarantee a higher level of privacy, as
long as one has enough friends who already are in the OSN.

In an interlinked community, an individual's privacy is a complex
property, where it is in constant mutual relationship with the
systemic properties and behavioral patterns of the community at
large. We provided quantitative insights into the dependence of an
individual's privacy to their respective community, and how far an OSN
provider can utilize this dependency to create shadow profiles.  Our
work does not improve the methods to create shadow profiles; we
limited ourselves to the application of existing methods to underline
an already existing risk. We showed that, as the network grows and its
members share their contact lists with the provider, the risk of
privacy leakage increases. Given the fact that this dependency is
present under generalized social interaction, we should consider
privacy as a collective concept, where individual privacy policies are
not sufficient to control private information.

\bibliographystyle{abbrv}

\end{document}